\documentclass{PoS}
\usepackage{calrsfs}
\usepackage{url}
\usepackage[square,comma,numbers]{natbib}

\renewcommand{\refname}{}

\title{First LHCb results from pA and Pb-Pb collisions}

\ShortTitle{First LHCb results from pA and Pb-Pb collisions}

\author{\speaker{L. Massacrier on behalf of the LHCb collaboration}\\
        LAL, Univ. Paris-Sud, CNRS/IN2P3, Universit\'e Paris-Saclay, Orsay, France \\
        IPNO, Univ. Paris-Sud, CNRS/IN2P3, Universit\'e Paris-Saclay, Orsay, France\\
        E-mail: \email{massacri@lal.in2p3.fr}}

\abstract{In 2015, the LHCb collaboration endorsed the proposal to pursue an ambitious heavy-ion physics program. In 2013, LHCb has demonstrated its capabilities to operate successfully in p-Pb and \linebreak Pb-p collisions, leading already to several important publications in the field. The measurements of the nuclear modification factor and forward-backward production of prompt and displaced J/$\psi$, $\psi$(2S) and $\Upsilon$(1S) states, as well as the production of prompt $D^{0}$ mesons, have allowed to extend the knowledge of Cold Nuclear Matter effects on open heavy flavours and quarkonium production. The measurement of Z-boson production, important to constrain nuclear PDFs, and the measurement of two-particle angular correlations, probing collective effects in the dense environment of high energy collisions, have also been performed. Furthermore, LHCb is the only experiment at the LHC that can be operated in fixed-target mode, owing to the injection of a small amount of gas inside the LHCb collision area. There have been several p-gas and Pb-gas data taking periods during Run 1 and beginning of Run 2. This fixed target programme is conducted at a center-of-mass energy of $\mathcal{O}$(100 GeV), and has great potential to bridge the gap of knowledge between SPS and RHIC in the domain of Quark Gluon Plasma physics. Finally, LHCb successfully participated to its first Pb-Pb data taking at the end of 2015. First studies show that up to semi-central Pb-Pb collisions can be successfully analysed to provide unique measurements in the forward region. In this proceeding, we will present a selection of LHCb results in pA collisions. We will also give some prospects for fixed target and Pb-Pb studies.}

\FullConference{Fourth Annual Large Hadron Collider Physics\\
		13-18 June 2016\\
		Lund, Sweden}

\begin{document}

\section{Introduction}

Ultra-relativistic heavy-ion collisions are used to study the nuclear matter at high temperature and pressure where the formation of the Quark Gluon Plasma (QGP) -- a state of matter which consists of asymptotically free quarks and gluons -- occurs for a very short amount of time. Then hadronization takes place as the fireball is cooling down. Heavy quarks are produced at the early stages of the collision only, and might interact with the deconfined medium, making them ideal probes of the QGP. It was indeed predicted that in hot nuclear matter quarkonia are suppressed due to color screening of the heavy quarks potential \cite{Matsui:1986dk}, and that the relative production probablities of quarkonia may provide information on the medium, in particular on its temperature. At LHC energies ($\mathcal{O}$(TeV)), the high charm quark density achieved may result in an enhanced probability to create charmonia from recombination of charm quarks during~\cite{Zhao:2011cv,Liu:2009nb,Thews:2000rj} or at the end~\cite{BraunMunzinger:2000px,Andronic:2011yq} of the deconfined phase, which makes the picture more complex. Open heavy flavours are also important to characterize the QGP properties (eg. transport coefficients). Indeed, heavy quarks can interact with the constituents of the medium and loose part of their energy via inelastic processes (gluon radiation) \cite{Gyulassy:1990ye,Baier:1996sk} or elastic scatterings (collisional processes) \cite{Thoma:1990fm,Braaten:1991jj,Braaten:1991we}. Quarkonia and open heavy flavour are also affected by Cold Nuclear Matter (CNM) effects, ie. effects related to the presence of a nuclei in the colliding system but without QGP formation. Proton-Nucleus collisions (pA or Ap), which are interesting by themselves, are therefore essential to interpret Nucleus-Nucleus data in order to disentangle QGP effects from CNM effects in such collisions. The main CNM effects affecting quarkonium production include, for instance, initial-state nuclear effects on the parton densities (shadowing) \cite{Albacete:2013ei}, the initial-state parton energy loss and final-state energy loss (coherent energy loss) \cite{Arleo:2012rs}, the final-state absorption of the pre-resonant heavy quark pair by the spectator nucleons (nuclear absorption) albeit small at LHC \cite{Ferreiro:2013pua} and the final-state interaction of the quarkonium with the produced medium (comovers) \cite{Ferreiro:2014bia}. Proton-nucleus collisions are also useful for the determination of nuclear parton distribution functions (nPDF) \cite{Salgado:2011wc}. The measurement of the Z electroweak boson in pA and Ap collisions in the LHCb acceptance permits to probe low $x_{A}$ values\footnote{$x_{A}$ is the momentum fraction of a certain parton inside a given nucleon bound in the nucleus A.}(2 $\times$ $10^{-4}$ - 3 $\times 10^{-3}$) in the forward case, and high $x_{A}$ value (0.2-1.0) in the backward case for an energy scale $Q^{2}$ = $M_{Z}^{2}$. Such a measurement (the first performed at the LHC in pA collisions) should offer large constraining power to nPDF fits especially at small $x_{A}$. LHCb, in fixed target mode, can also study CNM effects in p-Gas collisions and QGP physics in Pb-Gas\footnote{where Gas is a heavy enough nuclei (so far Argon (Ar)).} collisions in an energy domain $\mathcal{O}$(100 GeV), between SPS and RHIC energies, which remains unexplored so far. The fixed target program of LHCb will also provide valuable inputs for cosmic ray physics (eg. study of intrinsic charm, antiproton cross section measurement in pHe data). 

\section{The LHCb detector}

The LHCb detector is a single-arm forward spectrometer covering the pseudorapidity range \linebreak 2~$<~\eta~<$~5. It includes a high-precision tracking system (VELO) consisting of a silicon-strip vertex detector surrounding the pp interaction region, a large-area silicon-strip detector (TT) located upstream of a dipole magnet, and three stations of silicon-strip detectors and straw drift tubes (OT), placed downstream of the magnet. Different types of charged hadrons are distinguished using information from two ring-imaging Cherenkov detectors (RICH). Photons, electrons and hadrons are identified by a calorimeter system made of scintillating-pad (SPD), preshower detectors, an electromagnetic calorimeter (ECAL) and a hadronic calorimeter (HCAL). Muons are identified by a system composed of alternating layers of iron and multiwire proportional chambers. More details on the LHCb detector and its performances can be found in Refs. \cite{Alves:2008zz,Aaij:2014jba}.

\section{LHCb running mode and phase space coverage}

\begin{figure*}[htpb]
\centering
 \includegraphics[width=14.0cm,clip]{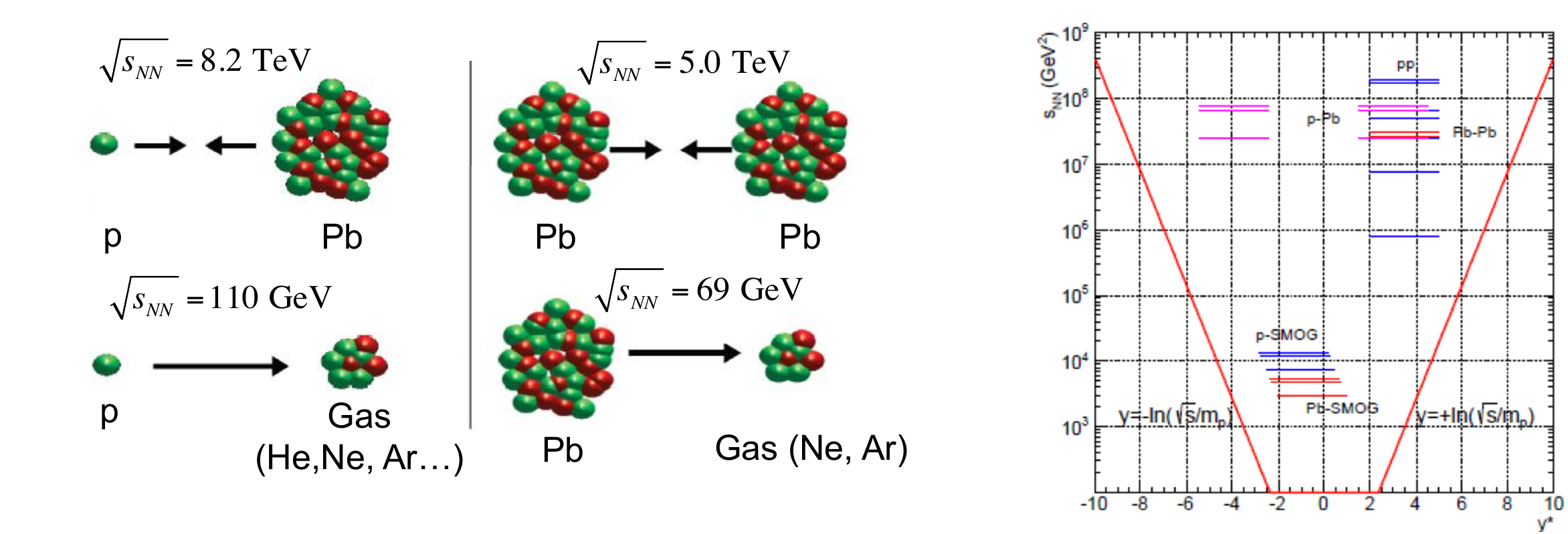}
\caption{Left: Cartoon of the various type of collisions explored by LHCb in collider mode (top) and fixed target mode (bottom). Right: Kinematic coverage in the nucleon-nucleon center-of-mass system of the LHCb detector for different beam-beam and beam-target combinations. The diagonal line indicates the beam rapidity. For asymmetric configuration, positive rapidity is in the direction of the proton or in case of fixed target operation, in the beam direction.}
\label{kine}       
\end{figure*}

LHCb is the only experiment at the LHC which can operate in parallel in collider mode and fixed target mode. The various type of collisions which can be studied in LHCb are shown in Fig. \ref{kine} left, for collider (top) and fixed target (bottom) modes. The nucleon-nucleon center-of-mass energy quoted corresponds to collisions with a 6.5 TeV proton beam and/or a 2.5 TeV lead beam. In Fig.~\ref{kine} right, the kinematic coverage of the LHCb detector is shown. A sizeable fraction of the space phase is covered by the detector, which compared to other experiments has the unique advantage to have precise tracking and vertexing, calorimetry and powerful particle identification in the full acceptance. In colliding beam mode, the forward/backward region is covered, while for fixed target running, the acceptance in the center-of-mass frame (CMS) is central to backward. Heavy nuclei collisions in fixed target mode would generate energy densities between those achieved at the SPS and those probed at RHIC. 

\section{Results from p-Pb and Pb-p collisions}

In 2013, LHCb collected pPb and Pbp data at a center-of-mass energy of about \linebreak $\sqrt{s_{NN}}$~=~5~TeV. In such asymmetric collisions, the nucleon-nucleon center-of-mass system is shifted by 0.47 unit of rapidity in the direction of the proton beam. In the forward (backward) configuration pPb (Pbp), the proton (lead) beam traverses LHCb from the vertex locator to the muon system, respectively. The LHCb acceptance is 1.5 $< \rm{y}_{CMS} <$ 4.0 in the forward configuration and \linebreak -5.0 $< \rm{y}_{CMS} <$ -2.5 in the backward configuration, leading to a common rapidy range coverage for both configurations of 2.5 $<\mid  \rm{y}_{CMS} \mid <$ 4.0. Analyses presented in this proceeding are performed with an integrated luminosity of $L_{int}$ = 1.1 nb$^{-1}$ for the forward configuration while in the backward configuration it amounts to $L_{int}$ = 0.5 nb$^{-1}$, except for the preliminary prompt D$^{0}$ analysis which was performed on a fraction of the available statistics ($L_{int}$ = 0.11 nb$^{-1}$ for the forward configuration and $L_{int}$ = 0.05 nb$^{-1}$ for the backward). Nuclear effects are usually quantified by the nuclear modification factor $R_{\rm{pA}}$ and the forward to backward ratio $R_{\rm{FB}}$. $R_{\rm{pA}}$ is defined as the production cross-section of a given particle in pA collisions divided by its production cross section in pp collisions at the same center-of-masse energy, and scaled by the atomic mass number A of the nuclei. While $R_{\rm{FB}}$ is the ratio of the production cross-section of a given particle in pA over its production cross-section in Ap configuration, measured in the same absolute center-of-mass rapidity range:
\begin{equation}
R_{\rm{pA}} = \frac{1}{A}\frac{\rm{d}^{2}\sigma_{\rm{pA}}(y,p_{T})/\rm{d}\sigma \rm{d}y}{\rm{d}^{2}\sigma_{\rm{pp}}(y,p_{T})/\rm{d}\sigma \rm{d}y}, \hspace{1 true cm} R_{\rm{FB}} = \frac{\sigma_{\rm{pA}}(+\mid y \mid, p_{\rm T})}{\sigma_{\rm{pA}}(-\mid y \mid, p_{\rm T})}.
\label{raa_eq}
\end{equation}
$R_{\rm{FB}}$ has the advantage of not relying on the pp reference cross-section and that part of the experimental systematic uncertainties and theoretical scale uncertainties cancel.

\subsection{Quarkonia production}

J/$\psi$, $\psi$(2S) and $\Upsilon$ are studied in the dimuon final state with a dimuon transverse momentum restricted to $p_{T} < 14$ GeV/$c$ for charmonia and to $p_{T} < 15$ GeV/$c$ for bottomonia. Prompt J/$\psi$ and $\psi$(2S) can be disentangled from J/$\psi$ and $\psi$(2S) coming from b-hadron decays thanks to the excellent vertexing capability of LHCb. The yields of prompt charmonia and charmonia from b-hadron decays are obtained, in each kinematic bin, from a simultaneous fit of the dimuon invariant mass and pseudo-proper time distributions. An unbinned extended maximum likelihood fit to the dimuon invariant mass of the $\Upsilon$ candidates was performed to determine the signal yields of $\Upsilon$(1S), $\Upsilon$(2S) and $\Upsilon$(3S). While the three $\Upsilon$ resonances are observed in the forward configuration, only $\Upsilon$(1S) gives a significant signal in the backward configuration. 
To determine the nuclear modification factor $R_{\rm{pPb}}$ of the J/$\psi$, $\psi$(2S) and $\Upsilon$(1S), their pp reference production cross-section at $\sqrt{s_{NN}}$ = 5 TeV are required (see Eq. \ref{raa_eq}). Since no data at this energy were available at the time of the measurement\footnote{A data dating period in pp collisions at $\sqrt{s}$ = 5 TeV has been completed end of November 2015 by the LHC.}, the reference cross-sections for J/$\psi$ and $\Upsilon$(1S) are obtained by a power-law fit to existing LHCb measurements at 2.76, 7 and 8 TeV\cite{ALICE:2013spa,LHCb:08082014sca}. To get the $\psi$(2S) reference pp cross-section at $\sqrt{s}$~=~5~TeV, the following assumption has been made:
\begin{equation}
\frac{\sigma_{\rm{pp}}^{J/\psi}(5 \hspace{0.1 true cm} \rm{TeV})}{\sigma_{\rm{pp}}^{\psi(2S)}(5 \hspace{0.1 true cm} \rm{TeV})} \approx \frac{\sigma_{\rm{pp}}^{J/\psi}(7 \hspace{0.1 true cm} \rm{TeV})}{\sigma_{\rm{pp}}^{\psi(2S)}(7 \hspace{0.1 true cm} \rm{TeV})},
\end{equation}
assuming that the systematic uncertainty on this hypothesis is negligible compared to the statistical uncertainty on the $\psi$(2S) measurement in pA collisions. The $\psi$(2S) and J/$\psi$ production cross-sections at $\sqrt{s}$~=~7~TeV are taken from the LHCb measurements in Refs \cite{Aaij:2011jh,Aaij:2012ag}. Under this assumption, the $\psi$(2S) nuclear modification factor can be derived from the J/$\psi$ nuclear modification factor\cite{Aaij:2013zxa}, with the following equation:
\begin{equation}
R_{\rm{pPb}}^{\psi(2S)} = \frac{\sigma_{\rm{pPb}}^{\psi(2S)}(5 \hspace{0.1 true cm} \rm{TeV})}{\sigma_{\rm{pPb}}^{J/\psi}(5 \hspace{0.1 true cm} \rm{TeV})} \frac{\sigma_{\rm{pp}}^{J/\psi}(5 \hspace{0.1 true cm} \rm{TeV})}{\sigma_{\rm{pp}}^{\psi(2S)}(5 \hspace{0.1 true cm} \rm{TeV})} \times R_{\rm{pPb}}^{J/\psi} \approx \frac{\sigma_{\rm{pPb}}^{\psi(2S)}(5 \hspace{0.1 true cm} \rm{TeV})}{\sigma_{\rm{pPb}}^{J/\psi}(5 \hspace{0.1 true cm} \rm{TeV})} \frac{\sigma_{\rm{pp}}^{J/\psi}(7 \hspace{0.1 true cm} \rm{TeV})}{\sigma_{\rm{pp}}^{\psi(2S)}(7 \hspace{0.1 true cm} \rm{TeV})} \times R_{\rm{pPb}}^{J/\psi}
\end{equation}

\begin{figure*}[htpb]
\centering
 \includegraphics[width=7.0cm,clip]{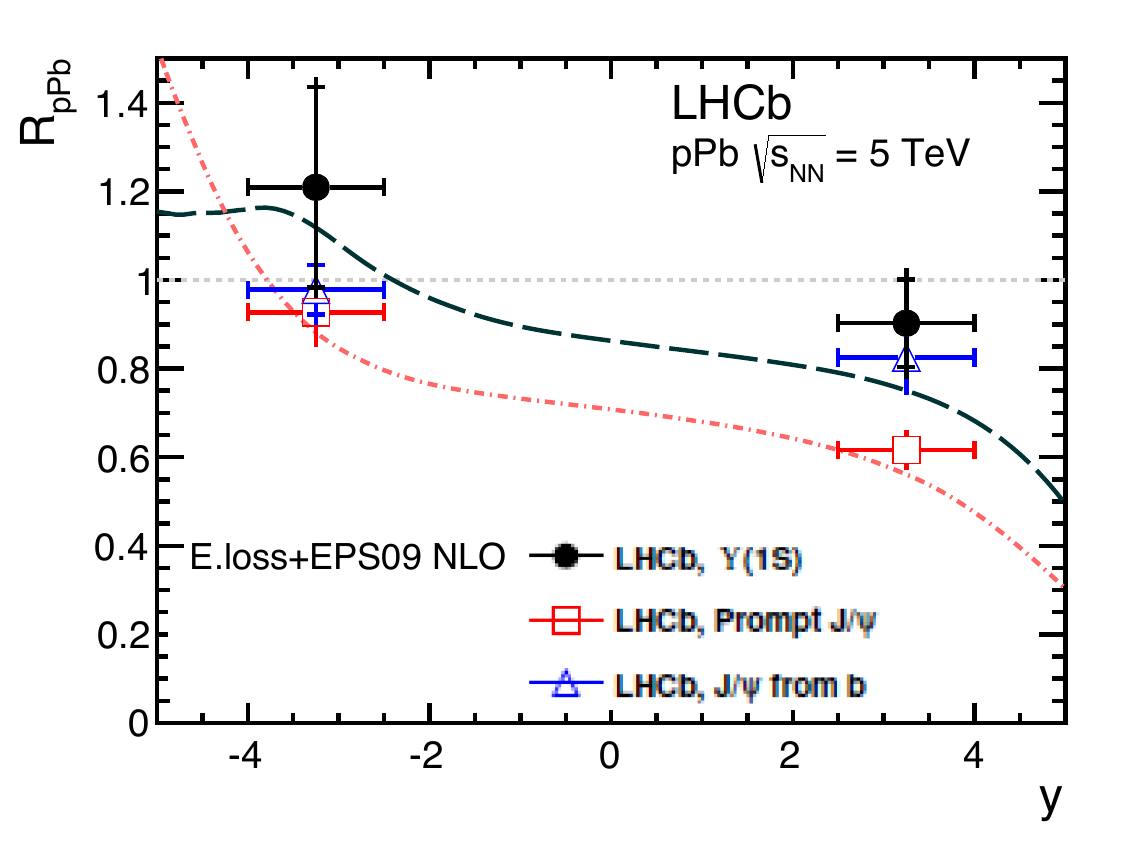}
 \includegraphics[width=7.0cm,clip]{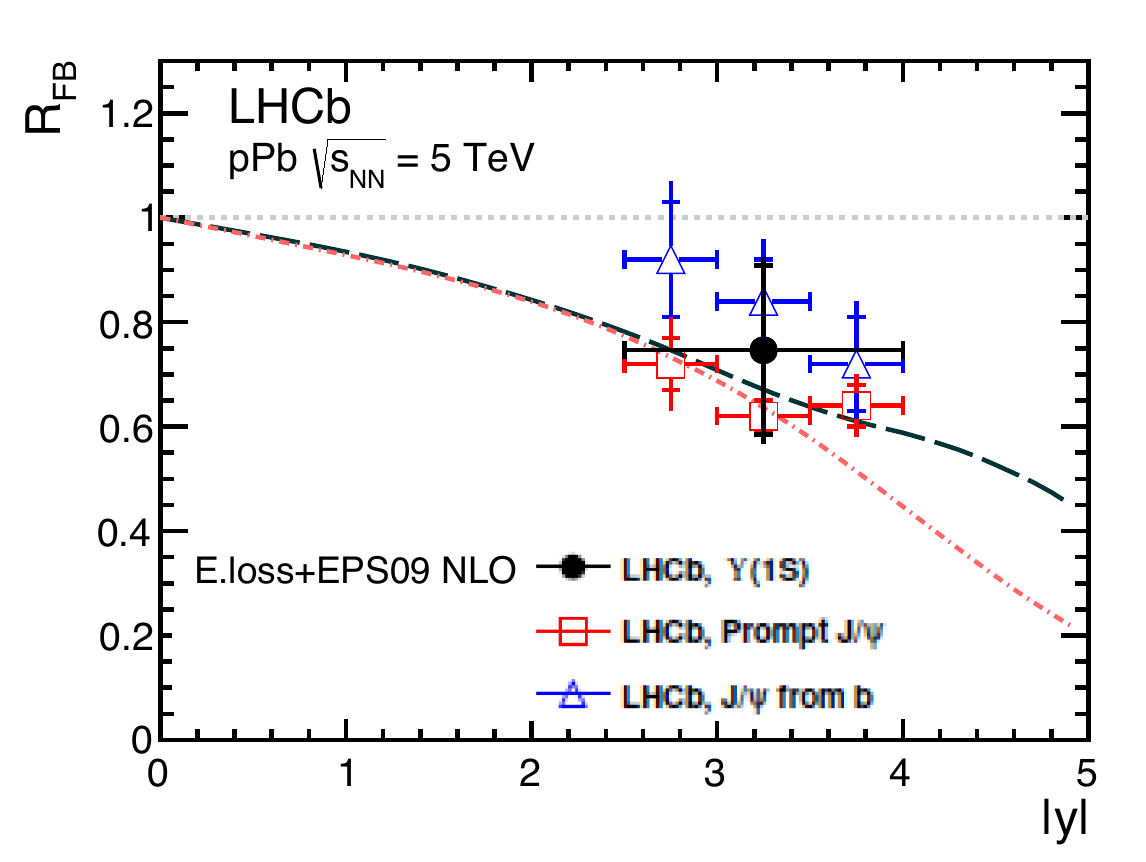}
\caption{$R_{\rm{pA}}$ (left) and $R_{\rm{FB}}$ (right) as a function of rapidity for prompt J/$\psi$ (red), J/$\psi$ from b (blue) and $\Upsilon$(1S) (black). The results are compared with theoretical predictions including energy loss and nuclear shadowing\cite{Arleo:2012rs}.}
\label{fig-3}       
\end{figure*}

Figure \ref{fig-2} left shows the $R_{\rm{pPb}}$ of prompt J/$\psi$ and $\psi$(2S) in the rapidity ranges \linebreak -4.0 $< y <$ -2.5 and 2.5 $< y <$ 4.0 \cite{Aaij:2013zxa,LHCb:psi2015}. The results are compared to several theoretical calculations of parton shadowing and coherent energy loss with or without shadowing\cite{Ferreiro:2013pua,delValle:2014wha,Albacete:2013ei,Arleo:2012rs}. A strong suppression of prompt J/$\psi$ is observed at forward rapidity which is compatible with most of the theoretical predictions. The comparison of prompt J/$\psi$ and $\psi$(2S) $R_{\rm{pPb}}$ suggests that $\psi$(2S) are more suppressed than J/$\psi$, especially in the backward region. Models describing the J/$\psi$ data might still be able to describe the forward $\psi$(2S) suppression but are not able to reproduce the suppression in the backward region. This intriguing result could be a first indication that another CNM effect is at play. Recently, a theoretical calculation based on comovers scenario \cite{Ferreiro:2014bia} tries to explain this behaviour. Figure \ref{fig-2} right shows the $R_{\rm{pPb}}$ of J/$\psi$ and $\psi$(2S) from b-hadrons in the rapidity ranges -4.0 $< y <$ -2.5 and 2.5 $< y <$ 4.0. J/$\psi$ from b are slightly suppressed in the forward region and both models including the shadowing effect can describe the data. In the backward region, J/$\psi$ from b are slightly less suppressed than prompt J/$\psi$ as expected from the models, however models are in worse agreement with the J/$\psi$ from b data. Given the large experimental uncertainties on the $\psi$(2S) from b measurement, no conclusions can be made on the comparison of the $\psi$(2S) from b suppression with respect to the J/$\psi$ from b suppression. Figure \ref{fig-3} left shows the measurement of $R_{\rm{pPb}}$ for $\Upsilon$(1S) as a function of rapidity \cite{Aaij:2014mza} compared with the $R_{\rm{pPb}}$ measurement of prompt J/$\psi$ and J/$\psi$ from b. $\Upsilon$(1S) is suppressed in the forward region while there is an indication for an enhancement of $\Upsilon$(1S) production with respect to pp in the backward region, which could be attributed to anti-shadowing. The $R_{\rm{pPb}}$ measurement of $\Upsilon(1S)$ agrees within uncertainties (albeit large in the backward region) with the $R_{\rm{pPb}}$ measurement of J/$\psi$ from b, reflecting the fact that similar CNM effects affect the b-hadrons production. $\Upsilon$(1S) data agree with coherent energy loss model including nuclear shadowing as parametrized with EPS09 \cite{Arleo:2012rs}. A comparison with more models is available in Ref. \cite{Aaij:2014mza}. Figure \ref{fig-3} right shows the forward-backward production ratio $R_{\rm{FB}}$ as a function of rapidity for $\Upsilon$(1S), prompt J/$\psi$ and J/$\psi$ from b. A smaller forward-backward asymmetry is observed for J/$\psi$ from b with respect to prompt J/$\psi$. The forward backward asymmetry measurement of $\Upsilon$(1S) and prompt J/$\psi$ agree with theoretical calculation of coherent energy loss including nuclear shadowing parametrized with EPS09\cite{Arleo:2012rs}.

\begin{figure*}[htpb]
\centering
 \includegraphics[width=7.0cm,clip]{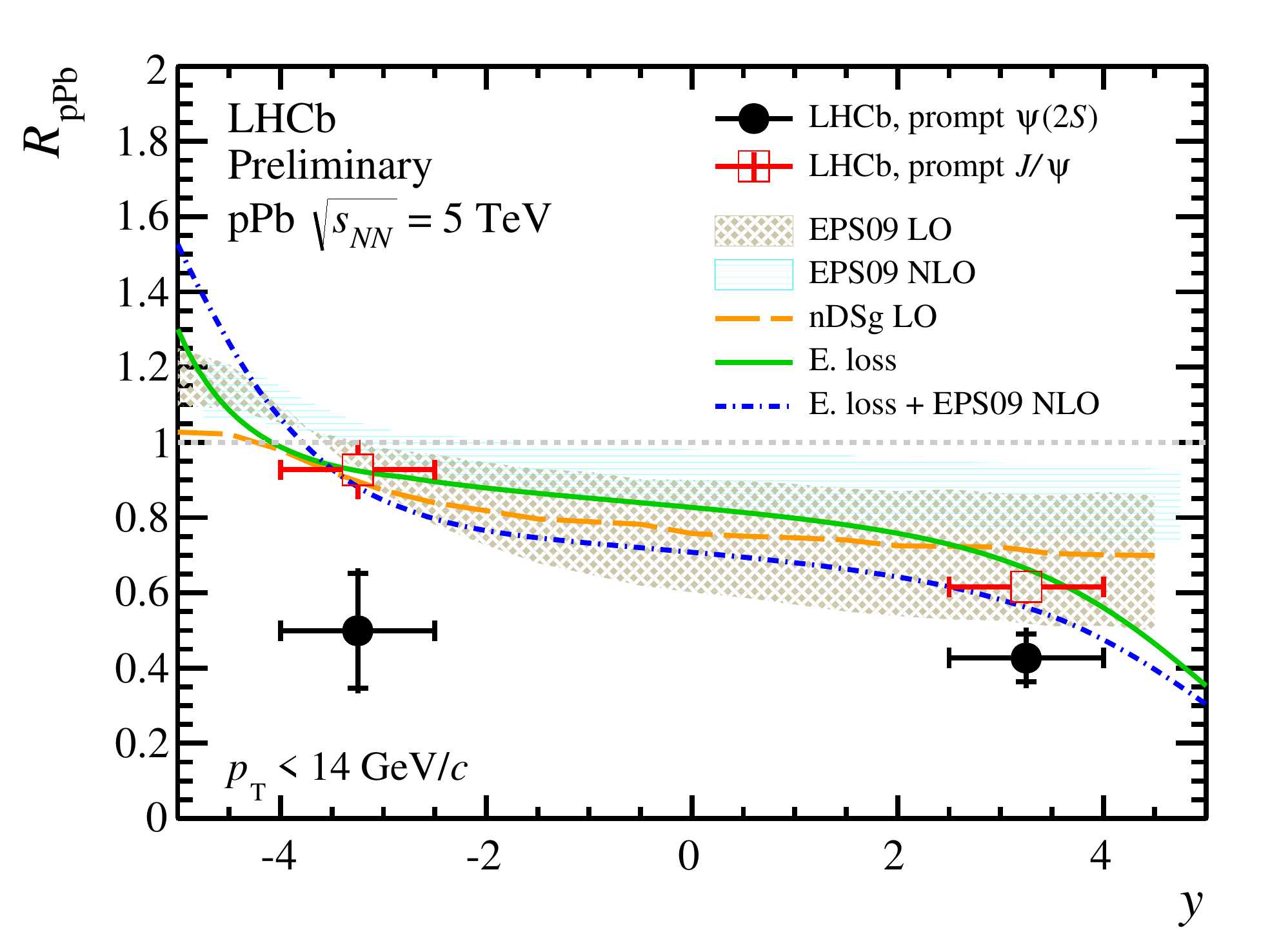}
 \includegraphics[width=7.0cm,clip]{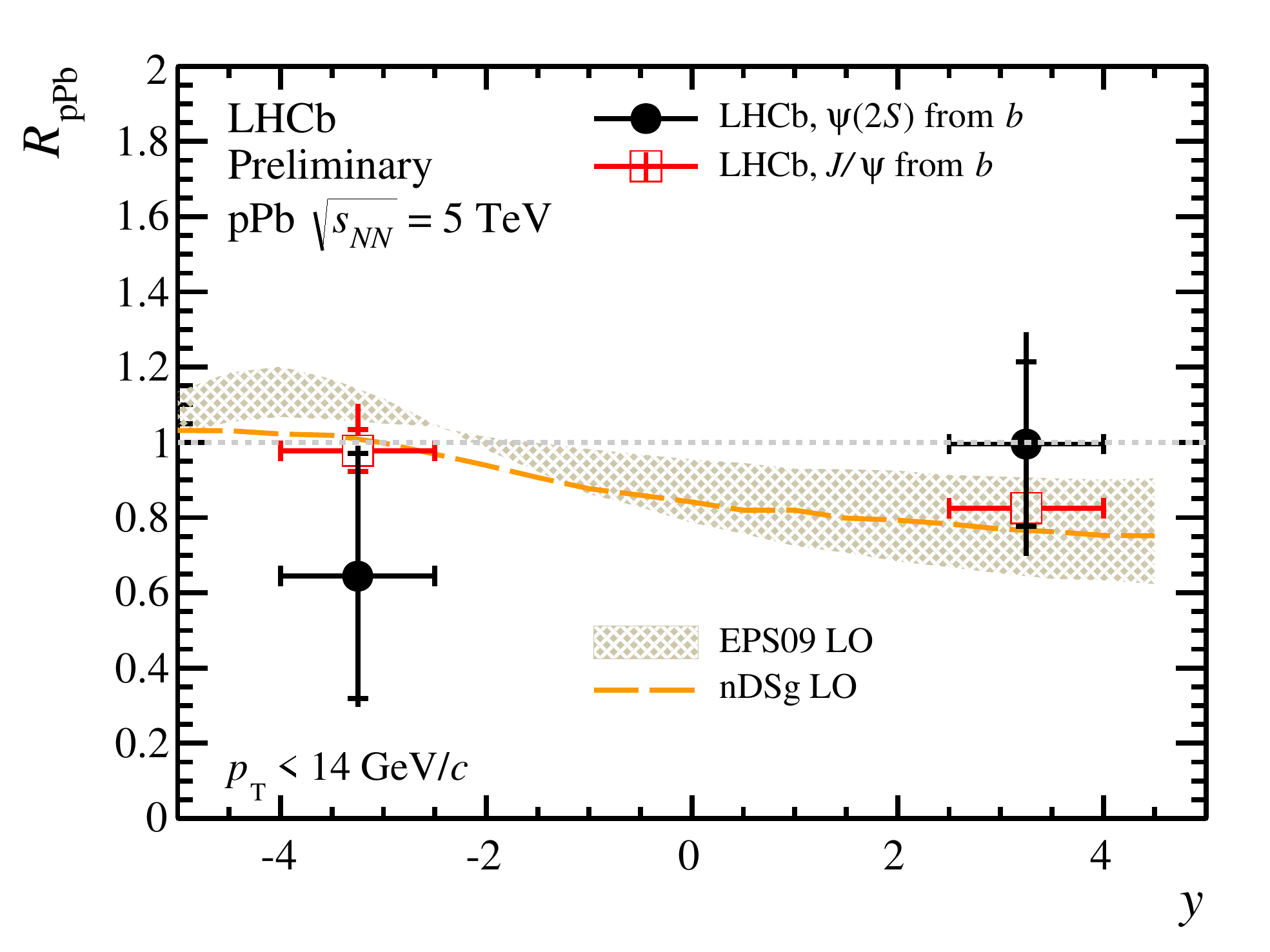}
\caption{Nuclear modification factor $R_{\rm{pPb}}$ as a function of rapidity for prompt J/$\psi$ and $\psi$(2S) (left) and for J/$\psi$ and $\psi$(2S) from b-hadrons (right). Results are compared with theoretical models from Refs. \cite{Ferreiro:2013pua,delValle:2014wha} (yellow dashed line and brown band), from Ref. \cite{Albacete:2013ei} (blue band) and from Ref. \cite{Arleo:2012rs} (green solid and blue dash-dotted lines). Only the models from \cite{Ferreiro:2013pua,delValle:2014wha} are available for $\psi$(2S) from b. } 
\label{fig-2}       
\end{figure*}

\subsection{Prompt $D^{0}$ production}

LHCb has the unique capability to measure prompt $D^{0}$ down to zero $p_{\rm T}$ in the forward region. The $D^{0}$ candidates are reconstructed in the $D^{0} \rightarrow K^{+} \pi^{-}$ and $D^{0} \rightarrow K^{-} \pi^{+}$ decay channels, in the kinematic range $p_{\rm T} <$ 8 GeV/c and for rapidities 1.5 $< y_{CMS} <$ 4 (forward case), \linebreak -5~$<~y_{CMS}~<$~-2.5 (backward case). $D^{0}$ candidates  are required to point to the primary vertex (PV) by using a requirement on the $\chi^{2}$ of the impact parameter ($\chi^{2}_{IP}(D^{0})$), defined as the difference in $\chi^{2}$ of a given PV computed with and without the $D^{0}$ candidate. Prompt $D^{0}$ signal is determined from an extended unbinned maximum likelihood fit to the M(K$\pi$) invariant mass coupled with a fit of the $log_{10}(\chi^{2}_{IP}(D^{0}))$ distribution which permits to obtain the fraction of $D^{0}$ from b-hadron decays. To obtain the prompt $D^{0}$ nuclear modification factor, the pp reference cross section at $\sqrt{s}$ = 5 TeV is obtained by extrapolating previous LHCb measurements at $\sqrt{s}$ = 7 TeV and 13 TeV\cite{Aaij:2013mga,Aaij:2015bpa}. 

\begin{figure*}[htpb]
\centering
 \includegraphics[width=7.0cm,clip]{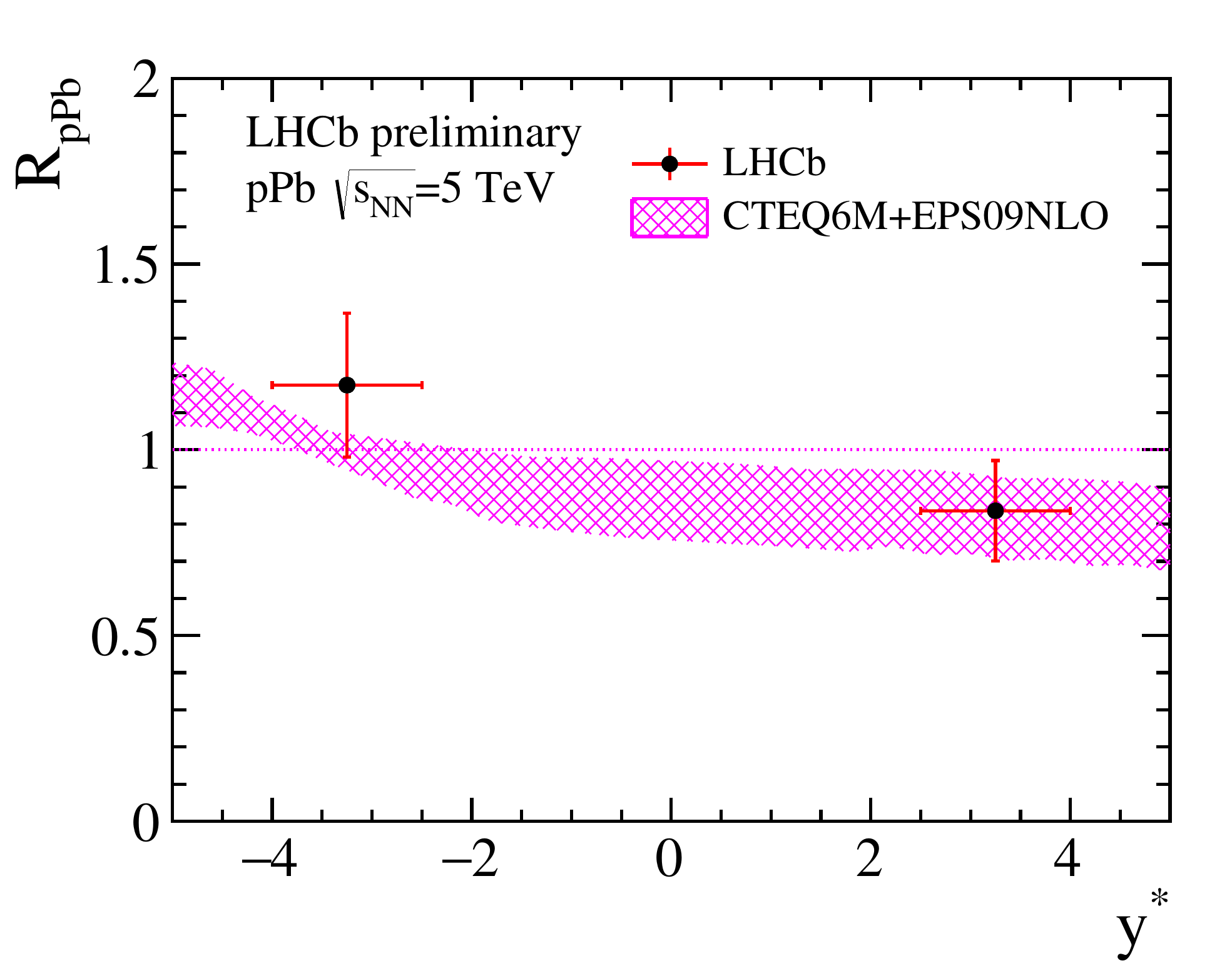}
 \includegraphics[width=7.0cm,clip]{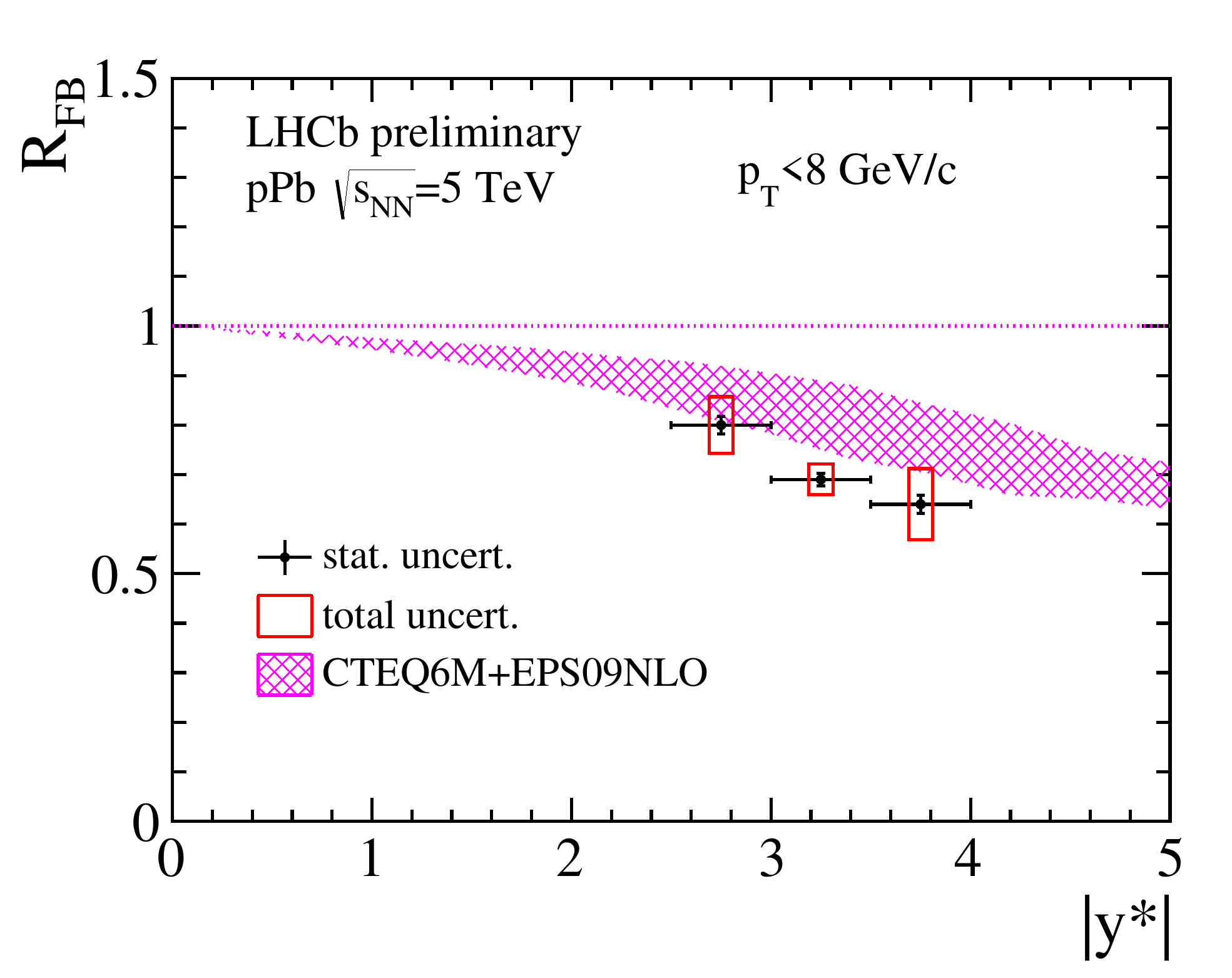}
\caption{Left: Nuclear modification factor $R_{pPb}$ (left) and Forward-Backward production ratio $R_{FB}$ (right) as a function of $y_{CMS}$ for prompt $D^{0}$ in the range $p_{\rm T} <$ 8 GeV/c. The total uncertainty is the quadratic sum of the statistical and systematic components.}
\label{pPb_D0}       
\end{figure*}

Figure \ref{pPb_D0} left shows the $R_{pPb}$ of prompt $D^{0}$ as a function of $y_{CMS}$ \cite{LHCb:2016huj}. The $R_{pPb}$ is measured to be smaller at forward rapidity than at backward rapidity. The $R_{pPb}$ as a function of $p_{\rm T}$ is also slightly smaller than unity at forward rapidity and doesn't exibit strong $p_{\rm T}$ dependence (not shown). Figure \ref{pPb_D0} right shows the $R_{FB}$ as a function of $y_{CMS}$. A clear forward-backward asymmetry is seen, suggesting significant CNM effects on prompt $D^{0}$ production. The asymmetry becomes stronger at larger rapidities. The $R_{FB}$ as a function of $p_{\rm T}$ also exhibits a clear forward-backward asymmetry with no strong $p_{\rm T}$ dependence (not shown). Both $R_{pPb}$ and $R_{FB}$ results are compared with next-to-leading order (NLO) prediction \cite{Mangano:1991jk} computed using CTEQ6M \cite{Stump:2003yu} and EPS09NLO \cite{Eskola:2009uj} parton distribution functions. In general, good agreement is found between LHCb measurements and the corresponding theoretical prediction.

\subsection{Z production}

Z boson candidates have been reconstructed in the dimuon decay channel, in the fiducial region defined by 2.0 < $\eta(\mu^{\pm})$ < 4.5, $p_{T}(\mu^{\pm})$ > 20 GeV/c and 60 $< m_{\mu^{+}\mu^{-}} <$  120 GeV/$c^{2}$. A clean signal of 11 candidates in the forward region and 4 candidates in the backward region has been observed\cite{Aaij:2014pvu}. The cross section for Z production is found to be $\sigma_{Z \rightarrow \mu^{+}\mu^{-}} = 13.5^{+5.4}_{-4.0} (\rm{stat.}) \pm 1.2 (\rm{syst.})$~nb in the forward direction and $\sigma_{Z \rightarrow \mu^{+}\mu^{-}} = 10.7^{+8.4}_{-5.1} (\rm{stat.}) \pm 1.0 (\rm{syst.})$~nb in the backward direction. The result in the forward region agrees with predictions from NNLO calculations using FEWZ \cite{Gavin:2010az} and the MSTW08 PDF set \cite{Martin:2009iq} with and without nuclear effects. In the backward region, data are higher than theoretical calculations. The forward-backward ratio has also been measured in the rapidity range 2.5 $< \mid y_{CMS} \mid <$ 4.0. It is found to be lower than expectations with a 2.2$\sigma$ deviation from one.
The present measurements have a limited statistical precision, preventing to put strong constraints on nuclear PDFs. 

\section{Prospects for fixed target studies}

The SMOG device (System to Measure the Overlap integral with Gas) allows for the direct injection in the LHC beampipe of noble gases at a pressure of the order of 1.5 $\times$ 10$^{-7}$ mbar, for fixed target studies. SMOG was initially developped to perform a precise determination of the luminosity with an uncertainty below 4\%. A pilot run of $p^{+}$ beam (Pb beam) on a Neon gas target was successfully performed in 2012 (2013) at a CMS energy of $\sqrt{s_{NN}}$ = 87 GeV (54 GeV). This first SMOG campaign was followed by several successful longer data taking periods in 2015: a p-Ne run at $\sqrt{s_{NN}}$ = 110 GeV ($\sim$ 12h), a p-He run at $\sqrt{s_{NN}}$ = 110 GeV ($\sim$ 8h), a p-Ar run at \linebreak $\sqrt{s_{NN}}$ = 110~GeV ($\sim$ 3 days), a p-Ar run at $\sqrt{s_{NN}}$ = 69 GeV (few hours) and a Pb-Ar run at \linebreak $\sqrt{s_{NN}}$ = 69 GeV (for about a week). In 2016, LHCb took p-He data at $\sqrt{s_{NN}}$ = 110 GeV for about 2 days. Preliminary studies of the p-Ne run ($\sqrt{s_{NN}}$ = 110 GeV) showed that clear J/$\psi$ and $D^{0}$ peaks are observed, with a good signal over background ratio (see \cite{LHCbPlots}). An integrated J/$\psi$ over $D^{0}$ ratio measurement should be at reach in the pNe sample, as well as differential measurements in the pAr ($\sqrt{s_{NN}}$ = 110 GeV) sample which contains larger statistics. Measurement of the antiproton production cross section in the various pHe samples is also ongoing. 

\section{Prospects for Pb-Pb studies}

At the end of 2015, LHCb sucessfully participated to its first Pb-Pb data taking. A first rough estimation of the integrated luminosity collected amounts to about 3-5 $\mu b^{-1}$. Data were collected with a minimum bias trigger, without any global event cut. The event reconstruction was then performed for event containing up to about 15000 clusters in the VELO detector, while events above this requirement are kept in order to have an unbiased centrality determination. \\
In heavy-ion collisions, centrality is a key quantity because it is related to the initial overlap region of the colliding nuclei. The size and the shape of the medium as well as the energy density depend on the geometry of the collision. The number of nucleons that participate in the collision is directly related to the collision geometry. Many quantities scale with the number of participating nucleons like for instance charged particle multiplicities. The centrality reach of LHCb is directly related to the LHCb tracking performances in a high multiplicity environment. 
A rough estimate of the centrality is ongoing using as input the energy deposition in the  electromagnetic calorimeter (Ecal) which should be an observable proportional to the centrality. The electromagnetic calorimeter (and hadronic calorimeter) have the advantages not to saturate also in most central collisions, contrary to the VELO. Centrality classes correspond to quantiles of the PbPb inelastic cross section. The energy distribution in Ecal can be fitted with a MC Glauber model \cite{Loizides:2014vua}, a geometrical model assuming that a Pb-Pb collision is an incoherent superposition of several nucleon-nucleon collisions.  The Ecal energy distribution is fitted with the function:
\begin{equation}
 f(E_{Ecal}) = C \times (f \times N_{part} + (1-f) \times N_{coll})
\end{equation}

\noindent{where $f$ and $C$ are free parameters of the fit. $N_{part}$ and $N_{coll}$ are the number of participating nucleons and number of binary collisions respectively. The fitted distribution is then divided into quantiles of Ecal energy (Ecal event-activity classes), which are already a good estimate of the collision centrality. Most peripheral collisions are not used in the fit since they are contaminated by soft QED and diffractive processes. The fitting range is 14 $< E_{Ecal} <$ 70 TeV. Figure \ref{fig-2} left shows the Ecal energy distribution divided into event-activity classes. The highest event-activity class corresponds to the class 0-10\% (ie. the most central collisions). Figure \ref{fig-2} right shows the correlation between the energy in the Ecal and the number of clusters in the VELO. There is a correlation between the number of VELO clusters and the Ecal total energy deposition for events containing up to 10000 clusters in the VELO. For events with large number of clusters in the VELO, the effect of the detector saturation can be seen. Preliminary studies based on Ecal energy observable demonstrated that LHCb centrality reach is about 50$\%$.


\begin{figure*}[htpb]
\centering
 \includegraphics[width=7.0cm,clip]{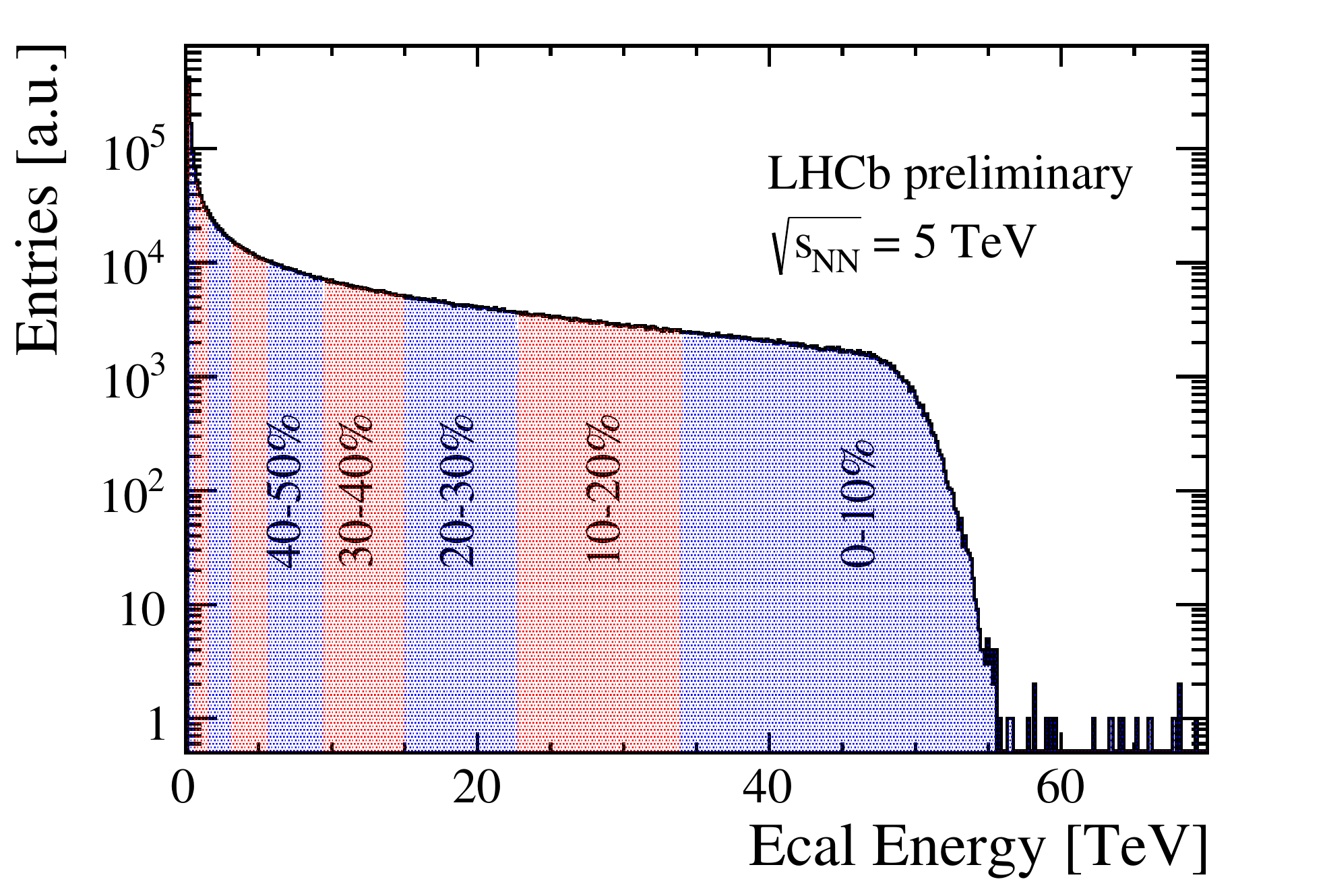}
 \includegraphics[width=7.0cm,clip]{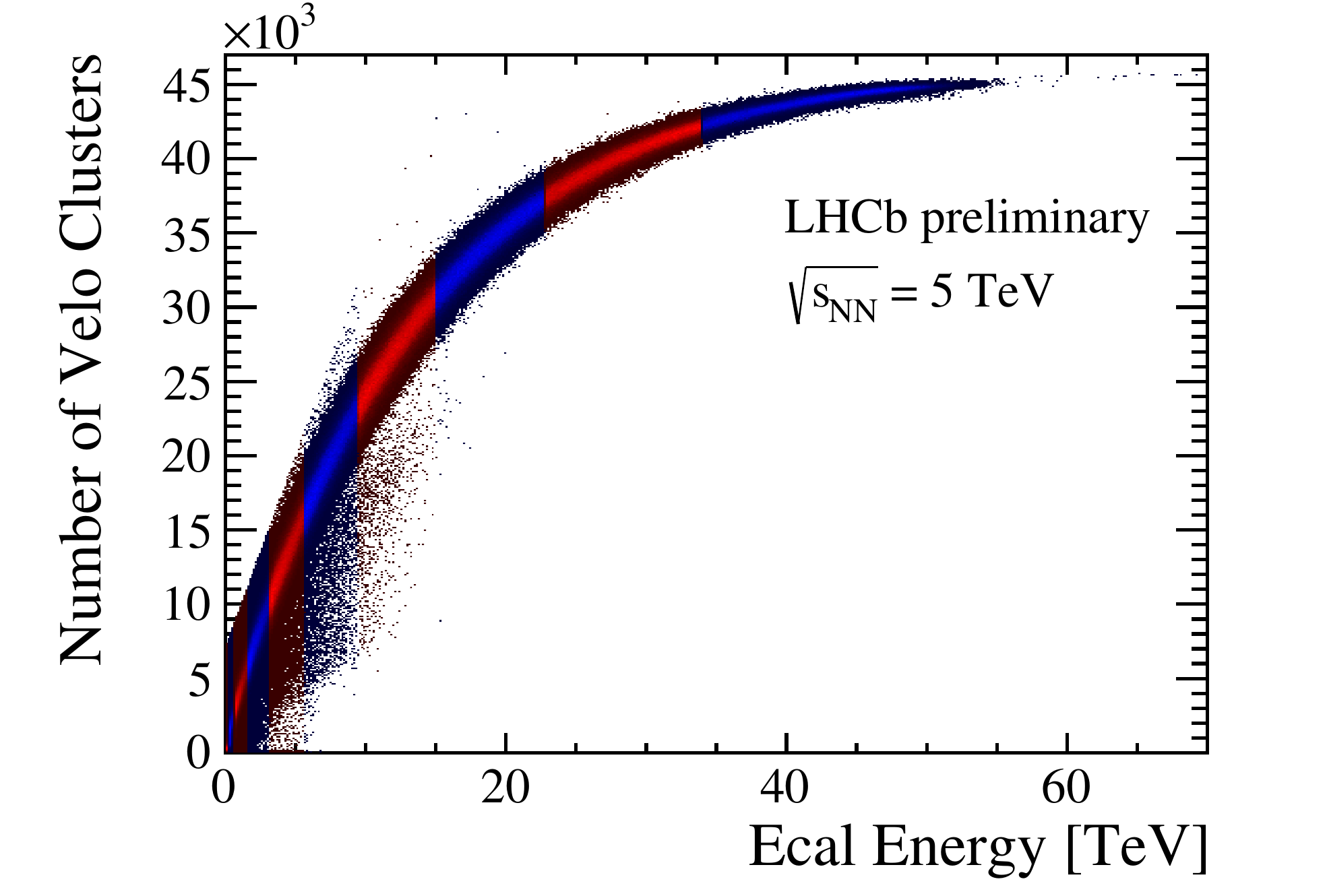}
\caption{Left: Ecal energy distribution in Pb-Pb collisions divided into event-activity classes. The highest event-activity class is the class 0-10\%. This class corresponds to most central collisions. Right: Correlation between the energy in the Ecal and the number of VELO clusters in Pb-Pb collisions.}
\label{fig-2}       
\end{figure*}



Figure \ref{fig-4} left shows the invariant mass distribution of the $D^{0}$ candidates, after standard selection cuts, in the Ecal event activity class 70-90$\%$. Figure \ref{fig-4} right shows similar invariant mass distribution for the Ecal event activity class 50-70$\%$. A clear signal can be seen in each event activity class. No signal is seen below the 50$\%$ event activity class, where the current limitations of the tracking algorithm are reached. The spectra are obtained using the full statistics collected.

\begin{figure*}[htpb]
\centering
 \includegraphics[width=7.0cm,clip]{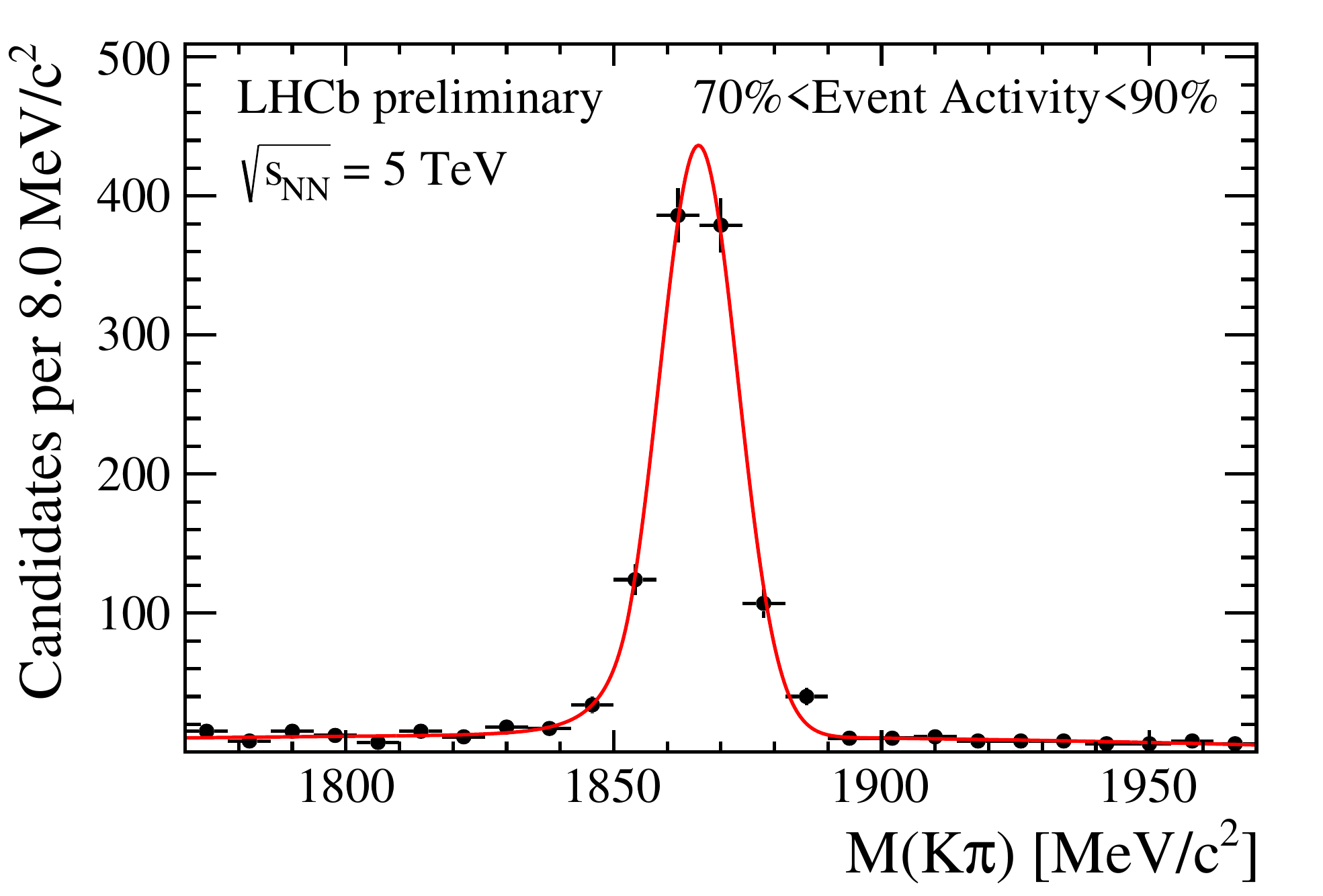}
 \includegraphics[width=7.0cm,clip]{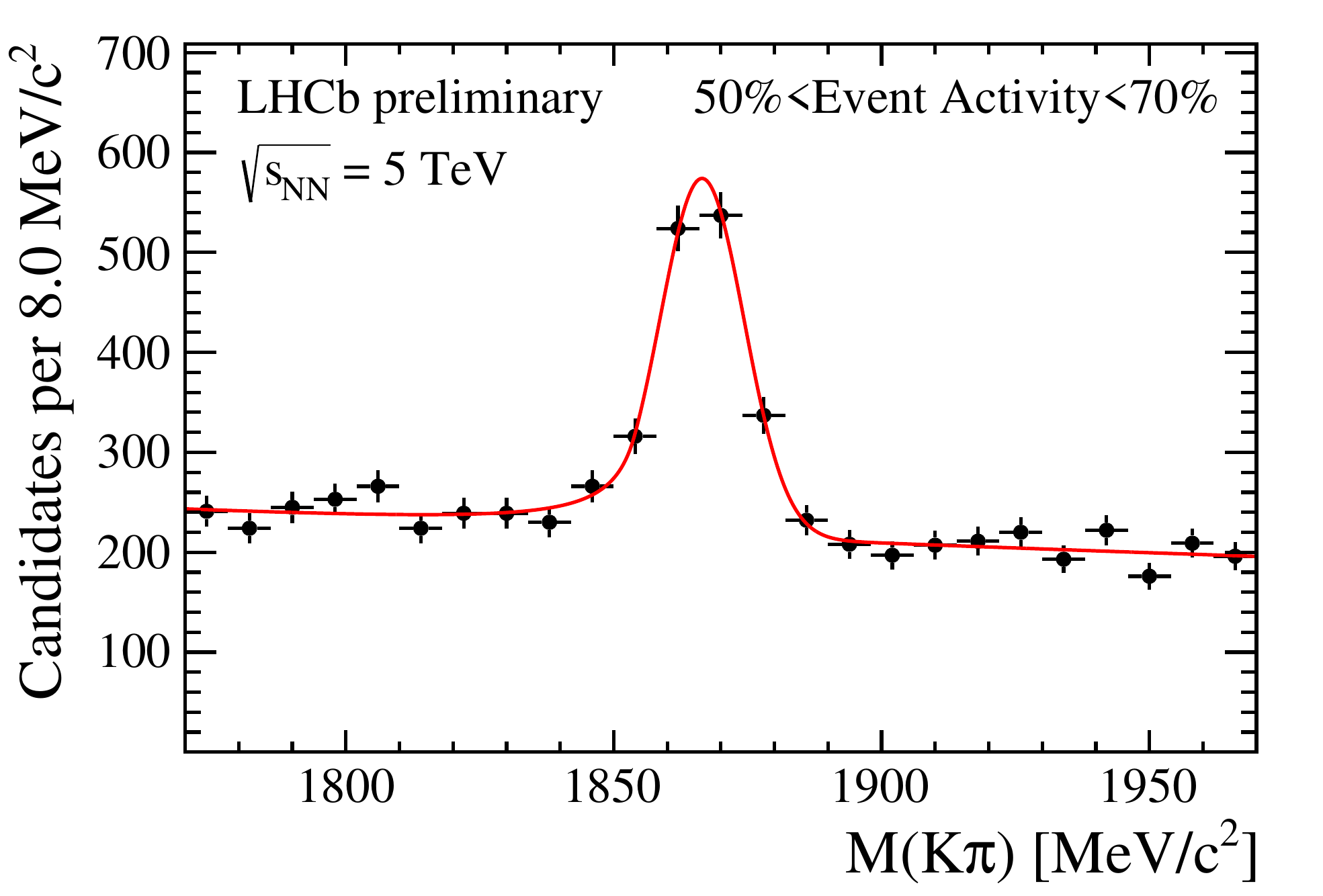}
\caption{Left: Invariant mass distribution of $D^{0}$ candidates in the event-activity class 70-90\%. Right: Invariant mass distribution of $D^{0}$ candidates in the event-activity class 50-70\% .}
\label{fig-4}       
\end{figure*}




Figure \ref{fig-7} left shows the J/$\psi$ signal obtained after standard selection cuts, in events containing two VELO tracks only. Figure \ref{fig-7} right shows the $p_{\rm T}^{2}$ distribution of these J/$\psi$ candidates in the invariant mass range 3050 $<$ M $<$ 3150 MeV/$c^{2}$. The signal is almost background free. The $p_{\rm T}^{2}$ distribution of the J/$\psi$ candidates follows a decreasing exponential law at low $p_{\rm T}^{2}$, as expected for coherently photoproduced J/$\psi$ in ultra-peripheral Pb-Pb collisions. These plots are performed using the full statistics collected. The analysis of ultra-peripheral collisions would benefit from the recently installed HERSCHEL detector, which was operational in the 2015 Pb-Pb data taking. This forward detector covering 5 $< |\eta| <$ 9 will permit to define rapidity gaps to select the events of interest.

\begin{figure*}[htpb]
\centering
 \includegraphics[width=7.0cm,clip]{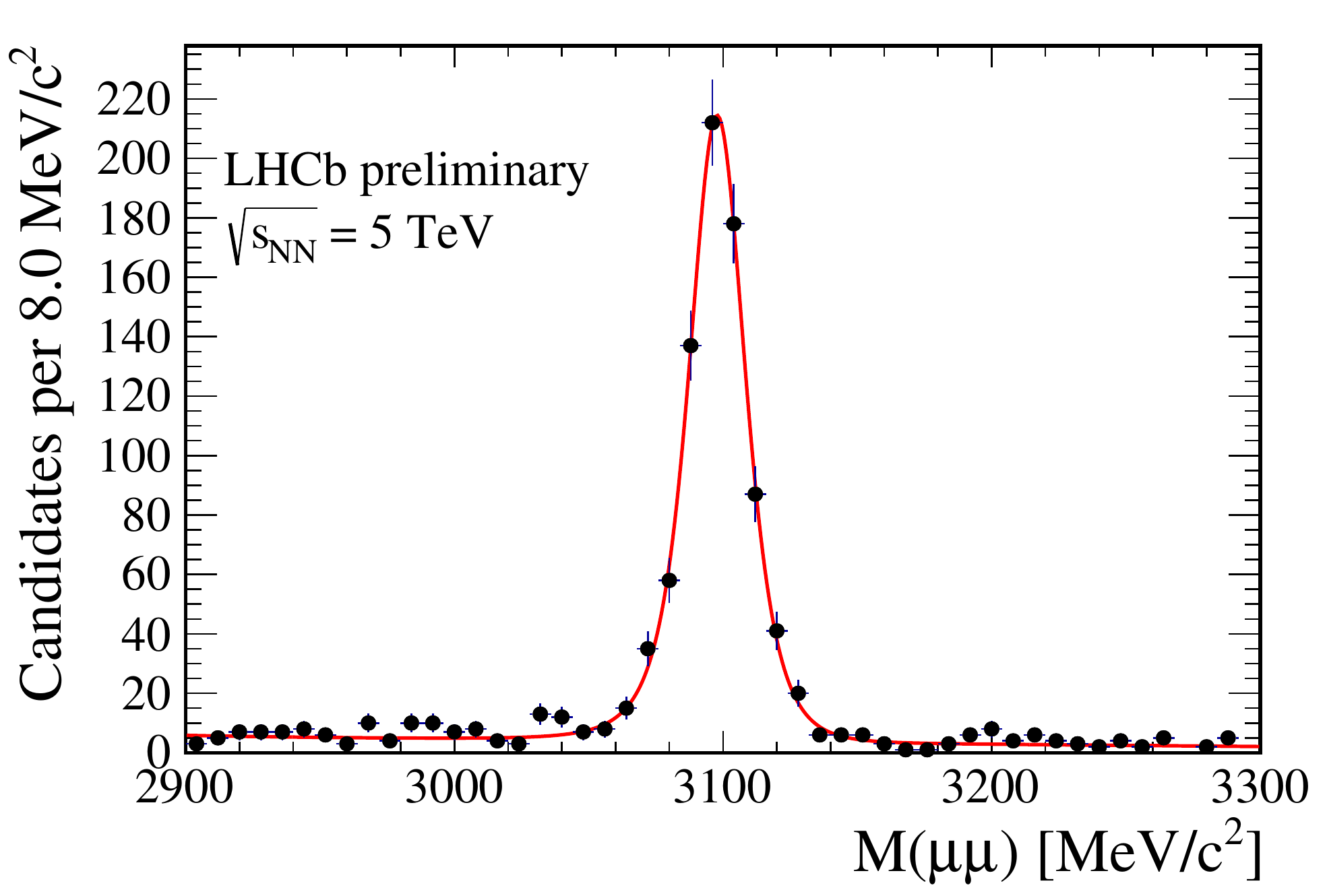}
 \includegraphics[width=7.0cm,clip]{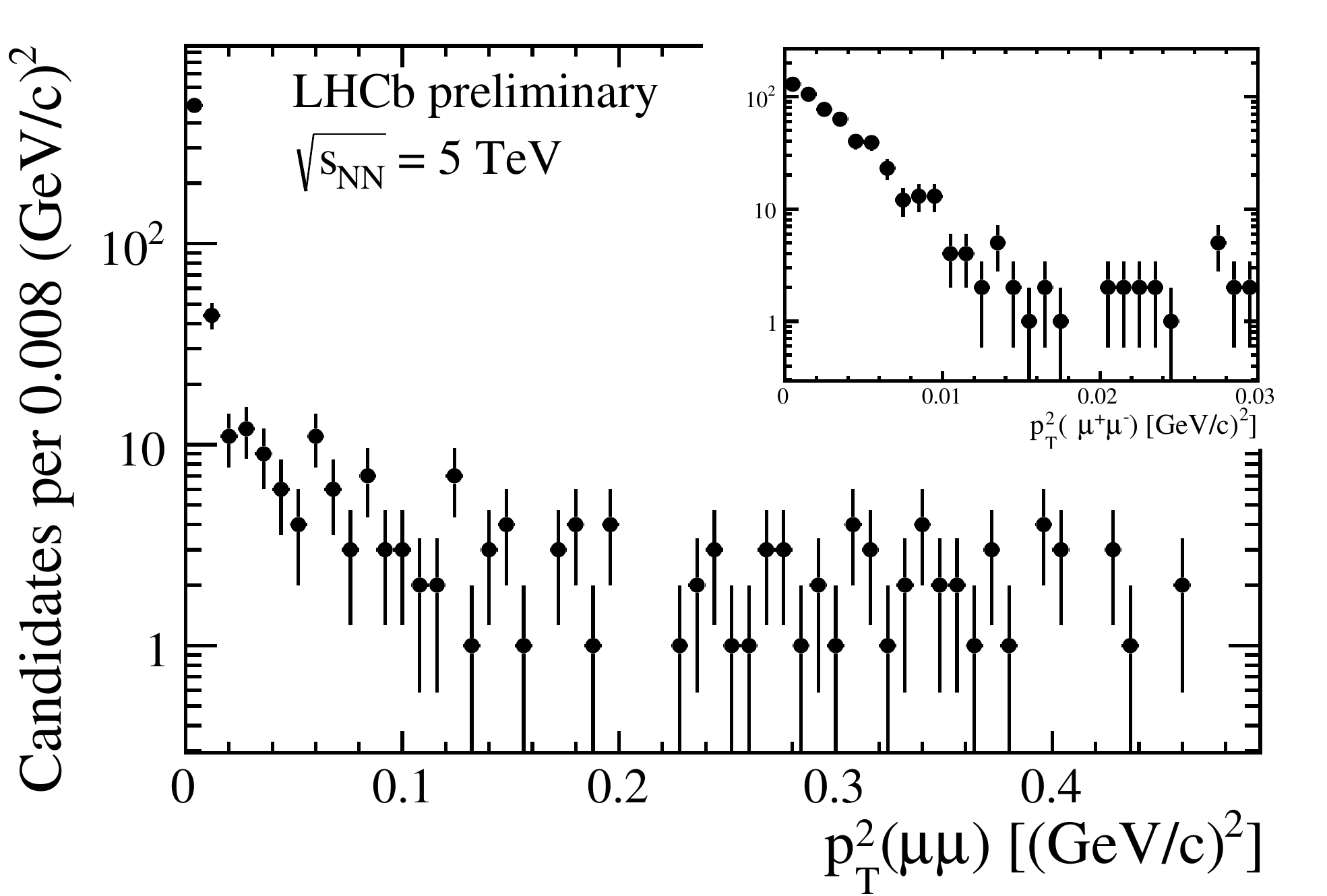}
\caption{Left: Invariant mass distribution of J/$\psi$ candidates passing the list of cut defined in the text. Right: $p_{\rm T}^{2}$ distribution of the dimuon candidates in the invariant mass range 3050 $<$ M $<$ 3150 MeV/$c^{2}$ passing the list of cuts defined in the text.}
\label{fig-7}       
\end{figure*}

\section{Conclusion} 

LHCb successfully participated in the p-Pb data taking in 2013 by measuring Cold Nuclear Matter effects on J/$\psi$, $\psi(2S)$, $\Upsilon(1S)$ and prompt $D^{0}$ production at $\sqrt{s_{NN}}$ = 5 TeV. J/$\psi$ from b, $\Upsilon$(1S) and prompt $D^{0}$ are less affected by CNM effects than prompt J/$\psi$ which exhibits a stronger suppression, especially in the forward region. Prompt $\psi$(2S) seems even more suppressed than prompt J/$\psi$ and the effect is more pronounced in the backward region. Models of coherent energy loss with and without nuclear shadowing effects give a good description of the data, except for the production of $\psi$(2S) in the backward region. The prompt $\psi$(2S) data suggest that another CNM effect could be at play. LHCb did also the first observation of forward Z production in proton-nucleus collisions at the LHC. The current analysis limited by statistical precision, will highly benefit from the larger statistics which will be collected end of 2016. Thanks to its SMOG system, LHCb is in unique position to do fixed target physics using collisions of proton or lead beams with various noble gas species (He, Ne, Ar). Thanks to this unique capability, LHCb will bridge the gap from SPS to LHC with a single apparatus. The LHCb detector has also collected Pb-Pb data for the first time at the end of 2015. Preliminary analysis of the data shows that LHCb will be able to analyse Pb-Pb collisions up to centrality of 50$\%$, allowing to conduct a rich physics program on heavy flavour, elecroweak, soft QCD and QGP physics. End of 2016, LHCb will participate in the p-Pb data taking at $\sqrt{s_{NN}}$ = 5 and 8 TeV. At 8 TeV, LHCb plans to collect an integrated luminosity of 10 $nb^{-1}$ per beam orientation. This will permit to improve the precision on the $\psi$(2S) and Z production measurements and will open the possibility to perform new measurements such as $\Upsilon$(3S), associated heavy flavour or Drell-Yan productions.

\paragraph{Aknowledgement:}
This work was supported by the French P2IO Excellence Laboratory.

\section{References}
\bibliographystyle{utphys}

\small
\bibliography{skeleton}


\end{document}